\documentclass[]{spie}  
\usepackage[]{graphicx}

\def\lsim{\mathbin{\lower 3pt\hbox
   {$\rlap{\raise 5pt\hbox{$\char'074$}}\mathchar"7218$}}} 
\def\gsim{\mathbin{\lower 3pt\hbox
   {$\rlap{\raise 5pt\hbox{$\char'076$}}\mathchar"7218$}}} 

\title{The Nuclear Spectroscopic Telescope Array (NuSTAR)} 

\author{Fiona A. Harrison\supit{a}, Steven Boggs\supit{b}, Finn Christensen\supit{c}, William Craig\supit{b,l}, Charles Hailey\supit{e}, 
Daniel Stern\supit{f}, William Zhang\supit{g},   Lorella Angelini\supit{g}, HongJun An\supit{e},
Varun Bhalereo\supit{a},  Nicolai Brejnholt\supit{c},  Lynn Cominsky\supit{h},  W. Rick Cook\supit{a}, Melania Doll\supit{e}, Paolo Giommi\supit{i},  \\ Brian Grefenstette\supit{a},  
Allan Hornstrup\supit{c},   Victoria M. Kaspi\supit{j}, Yunjin Kim\supit{f},  
Takao Kitaguchi\supit{a},  Jason Koglin\supit{e},  Carl Christian Liebe\supit{f},  Greg Madejski\supit{k}, Kristin Kruse Madsen\supit{a}, Peter Mao\supit{a},  David Meier\supit{f}, 
Hiromasa Miyasaka\supit{a},  Kaya Mori\supit{e},  
Matteo Perri\supit{i},  Michael Pivovaroff\supit{l}, \\ Simonetta Puccetti\supit{i},
Vikram Rana\supit{a},  Andreas Zoglauer\supit{b}
\skiplinehalf
\supit{a} Caltech Division of Physics, Mathematics and Astronomy, Pasadena, USA; \\
\supit{b} U.C. Berkeley Space Sciences Laboratory, Berkeley, CA; \\
\supit{c}   National Space Institute, Technical University of Denmark, Copenhagen, DK  ; \\
\supit{e} Columbia University, NY, USA; \\
\supit{f}  Jet Propulsion Laboratory, Pasadena, CA, USA; \\
\supit{g} Goddard Space Flight Center, Greenbelt, MD, USA; \\
\supit{h} Sonoma State University, Sonoma, CA, USA; \\
\supit{i} ASI Science Data Center, Rome, IT ; \\
\supit{j} McGill University Department of Physics, Montreal, Canada; \\
\supit{k} Stanford University, SLAC, USA; \\
\supit{l} Lawrence Livermore National Laboratory, Livermore, CA, USA; \\
}

\authorinfo{Further author information: (Send correspondence to F.A.H.) F.A.H.: E-mail: fiona@srl.caltech.edu}

\begin{document} 
  \maketitle 

\begin{abstract}
The {\em Nuclear Spectroscopic Telescope Array (NuSTAR)} is a NASA Small Explorer mission that will carry the first focusing hard X-ray (5 -- 80 keV)
telescope to orbit.   {\em NuSTAR} will offer a factor 50 -- 100 sensitivity improvement compared to previous collimated or coded mask imagers that 
have operated in this energy band.  In addition, {\em NuSTAR} provides sub-arcminute imaging with good spectral resolution over a 12-arcminute
field of view.   After launch, {\em NuSTAR} will carry out a two-year primary science mission that focuses on four key
programs: studying the evolution of massive black holes through surveys carried out in fields with excellent multiwavelength 
coverage,  understanding the population of compact objects and the nature of the massive black hole in the center of the Milky Way,
constraining explosion dynamics and nucleosynthesis in supernovae, and probing the nature of particle acceleration in relativistic
jets in active galactic nuclei.    A  number of additional observations will be included in the primary mission, and a guest observer program
will be proposed for an extended mission to expand the range of scientific targets.    The payload consists of two co-aligned depth-graded
multilayer coated grazing incidence optics focused onto solid state CdZnTe pixel detectors.   To be launched in early 2012 on a Pegasus rocket
into a low-inclination Earth orbit,  {\em NuSTAR} largely avoids SAA passages, and will therefore have low and stable detector backgrounds.  
The telescope achieves a 10.15-meter focal length through on-orbit deployment of an extendable mast.
An aspect and alignment metrology system enable reconstruction of the absolute aspect and variations in the telescope alignment
resulting from mast flexure during ground data processing.  Data will be publicly available at GSFC's High Energy Astrophysics Science
Archive Research Center (HEASARC) following validation at the science operations center located at Caltech.
\end{abstract}



\section{INTRODUCTION}
\label{sec:intro}  

The last decade has seen a major technological advance in hard X-ray/soft gamma-ray astronomy -- the ability to focus efficiently -- 
enabling instruments that improve sensitivity by orders of magnitude compared to the collimators and coded-mask cameras previously
used to observe the cosmos at these energies.   Focusing instruments achieve large concentration factors, such that their collecting
area is significantly larger (by factors of 1000 or greater) than the detector area used to register the signal.   In the hard X-ray
band, where particle interactions result in high detector backgrounds, large concentration factors result in enormous improvements in the signal to background
ratio over coded mask cameras, where telescope effective areas are typically less than ($\sim$50\%) of the detector area.   
Focusing telescopes operating at energies above 10~keV have recently been developed and deployed on balloon platforms\cite{hcc+05,raa+02,tko+05},
and will be incorporated on two approved space experiments; 
{\em The Nuclear Spectroscopic Telescope Array (NuSTAR)} NASA Small Explorer, and the JAXA {\em Astro-H}\cite{tak+10}
mission, as well as  the proposed {\em New Hard X-ray Mission (NHXM)}\cite{pta+09}.

Extending focusing to the hard X-ray band requires a combination of low graze-angle and/or depth-graded multilayer-coated optics combined with
imaging detectors utilizing high-atomic-number materials.     For grazing-incidence X-ray optics,  the graze angle (angle at which X-rays are
reflected from a shell) range for which 
efficient reflectance can be achieved scales approximately inversely with energy.   To achieve high-energy (10 -- 60 keV) response with traditional
metal coatings the optics design must utilize graze angles of a few arcminutes or less, which requires small-radius, tightly-nested optics
shells.  Since the field of view of a
grazing incidence telescope is approximately equal to the average graze angle,  metal-coated optics will have small fields of view at
high energy.   To overcome this limitation, a number of future astronomical telescopes will employ depth-graded multilayer coatings\cite{chw+92}, 
which exploit
the principal of Bragg reflection to increase the graze angles at which significant reflectance can be achieved.    This enables high
throughput with moderate ($\sim10 ^\prime$ ) FoV.     

The {\em NuSTAR} Small Explorer mission will be the first astronomical telescope on-orbit to utilize the new generation of hard X-ray
optics and detector technologies to carry out high-sensitivity observations at X-ray energies significantly greater than 10~keV.  
{\em NuSTAR}, based in large part on the technologies developed for the {\em High-Energy Focusing Telescope (HEFT)}\cite{hcc+05} balloon
experiment, was selected after a competitive Phase A study for implementation, and the mission is now in Phase D with a launch scheduled for the first part
of calendar year 2012.   This paper describes {\em NuSTAR}'s  two-year primary science program, the implementation of the science
instrument, the mission design, and plans for data distribution and archiving.   

\begin{figure}
\begin{center}
\begin{tabular}{c}
\includegraphics[height=6.5cm]{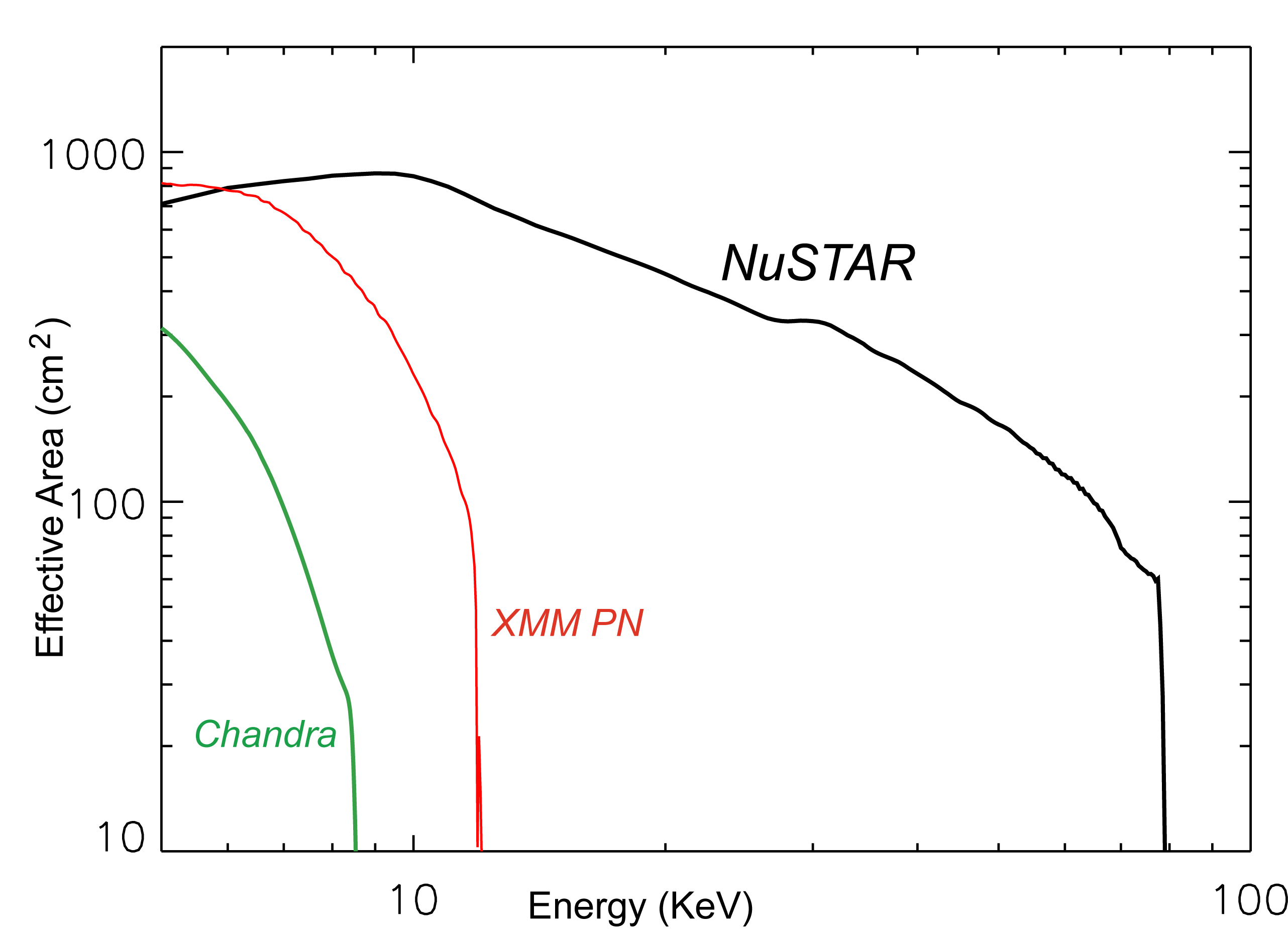}
\end{tabular}
\end{center}
\caption[area] 
{\label{fig:area} 
Effective area for two telescopes as a function of energy compared with the {\em Chandra} and {\em XMM} focusing telescopes.  {\em NuSTAR} utilizes
a low graze angle design combined with depth-graded multilayer coatings to extend sensitivity to 80 keV.  Focusing in this band provides improvements
in sensitivity by a factor 50 - 100 compared to hard X-ray collimated or coded mask experiments with larger collecting area}
\end{figure}

\section{SCIENTIFIC PERFORMANCE} 

{\em NuSTAR} will fly two co-aligned focusing hard X-ray telescopes consisting of multilayer-coated, grazing-incidence optics and
shielded solid state CdZnTe pixel detectors with a 10.15-meter focal length.   Figure~\ref{fig:area} shows the total effective area for
both telescopes as a function of energy, with a comparison to {\em Chandra} and {\em XMM}.   The energy band extends from
about 5~keV to 80~keV, being limited at the low-energy end by the optics thermal cover and shield entrance
window, and at the high energy end by the K-edge (at 78.4 keV) in the Platinum mirror coatings.    

\begin{figure}
\begin{center}
\begin{tabular}{c}
\includegraphics[height=7.5cm]{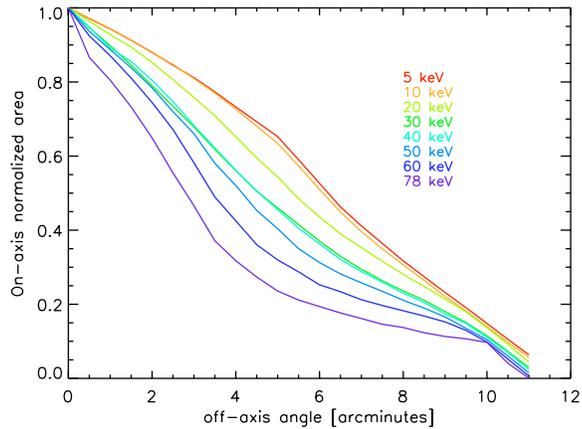}
\end{tabular}
\end{center}
\caption[offaxis] 
{\label{fig:offaxis} 
Effective area as a function of off-axis angle, as a fraction of on-axis area, for several energies.}
\end{figure}

Table~\ref{tab:perf} provides the current best estimates of the
key instrument performance parameters.  The 45$^{\prime\prime}$ angular resolution is projected based on mechanical
metrology of the first flight optic, which is currently more than 75\% complete, combined with a model of contributions
from on-orbit thermal distortions and aspect reconstruction.  The field of view is
energy-dependent due to changes in multilayer reflectance as a function of energy and optics shell radius, which results in overall
loss of reflectance and more vignetting at high energy (see Figure~\ref{fig:offaxis}).   
The spectral resolution is 500~eV at energies below $\sim$30~keV, and increases to 1.2~keV at the upper end of the energy range.    
The 2~$\mu$sec temporal resolution, determined by the
bit rate allocated in the telemetry stream for time tags, is more than adequate to meet scientific requirements.  The intrinsic temporal resolution of the detector is better than 1~$\mu$sec.   The target of opportunity (ToO) response time is required to be
less than 24 hours, however, on average the turnaround will be faster, with targets typically acquired within 6~hours.

\begin{table}
\caption{Key instrument performance parameters.}
\label{tab:perf}   
\begin{center}
\begin{tabular}{|l|c|}
\hline
Energy range & 5 -- 80 keV \\ \hline
Angular resolution (HPD)  & 45$^{\prime\prime}$  \\ \hline
Angular resolution (FWHM) & 9.5$^{\prime\prime}$ \\ \hline
FoV (50\% resp.)  at 10 keV &  10$^\prime$ \\ \hline
FoV  (50\% resp.) at 68 keV & 6$^\prime$  \\ \hline
Sensitivity (6 - 10 keV) [10$^6$ s, 3$\sigma$, $\Delta$E/E = 0.5] & $2 \times 10^{-15}$ erg/cm$^2$/s \\ \hline
Sensitivity (10 - 30 keV) [10$^6$ s, 3$\sigma$, $\Delta$E/E = 0.5] & $1 \times 10^{-14}$ erg/cm$^2$/s \\ \hline
Background in HPD (10 - 30  keV) & $6.8 \times 10^{-4}$ cts/s \\ \hline
Background in HPD (30 -- 60 keV) & $4.0 \times 10^{-4}$ cts/s  \\ \hline
Spectral resolution (FWHM) &  500~eV at 10~keV, 1.2 keV at 80 keV \\ \hline
Strong source ($> 10\sigma$) positioning &  1.5$^{\prime\prime}$ (1-$\sigma$) \\ \hline
Temporal resolution & 2 $\mu$sec \\ \hline
Target of Opportunity response & $< 24$ hours \\ \hline
\end{tabular}
\end{center}
\end{table} 

\section{BASELINE MISSION SCIENCE PLAN} 

\begin{table}[h]
\caption{Core science program.}
\label{tab:keyscience}
\begin{center}
\begin{tabular}{|l|l|}
\hline
{\bf Key science goal } & {\bf Observations}  \\
\hline\hline
Locate massive black holes & Deep and wide-field surveys  \\ 
                                                  & in GOODS, COSMOS, XBootes \\ \hline
Study the population of compact objects in the Galaxy & Galactic survey centered on Sgr A* \\ \hline
Understand explosion dynamics and nucleosynthesis in  & Pointed observations of Cas A, SN1987A, Tycho \\ 
core collapse and Type Ia SNe & ToO observations of Ia SNe \\ \hline
Constrain particle acceleration in relativistic jets  & Contemporaneous multiwavelength observations  \\ 
in supermassive black holes & of GeV and TeV blazars \\ \hline
\end{tabular}
\end{center}
\end{table}

The two-year {\em NuSTAR} baseline science mission begins after a 1-month on-orbit checkout period.   During the baseline mission
{\em NuSTAR} will focus on a set of four key objectives (Table~\ref{tab:keyscience}).  If the currently estimated performance is
achieved on-orbit, completing these objectives will require about eighteen months of observation time.   
The remaining six months will be used to perform additional targeted programs.    Although survey fields and specific sources
for the key programs have been identified, the detailed designs of the observations are not yet complete.  The science team is also in 
the process of prioritizing the additional science objectives.    The final baseline science observing plan will be determined six 
months prior to launch, although
alterations may be made after launch as a result of measured on-orbit performance or preliminary data analysis.

Table~\ref{tab:keyscience} summarizes the key objectives and the associated observations.    The primary extragalactic
science objective is to understand the physical processes that drive the evolution of supermassive black holes and galaxies
at redshift $z \lsim 1$.    {\em NuSTAR's} specific contribution will be to measure the evolution of obscured active galactic
nuclei (AGN),  study the characteristics of the  galaxies that host them, and determine
if they evolve similarly to the well studied unobscured population.
In addition, {\em NuSTAR} will undertake simultaneous observations of blazar AGN with {\em Fermi} and the TeV gamma-ray
telescopes  Veritas and HESS, as well as with ground-based optical and radio telescopes in order to study particle
acceleration in jets.     In the Galaxy, {\em NuSTAR} will advance the understanding of explosion dynamics and nucleosynthesis
in supernovae by mapping Cas~A in the radioactive decay of $^{44}$Ti (68~keV).   Other remnants that will be studied include 
SN~1987A, Tycho, and G1.9+03.  {\em NuSTAR} will also survey regions of the Galaxy, with the primary survey field being
the few square degrees
centered on Sgr~A*, a region that contains 1\% of the stellar mass, but 10\% of the massive young stars.   The Galactic surveys will
identify hundreds of compact stellar remnants even in obscured regions, and will map diffuse features associated with
molecular cloud complexes.    The {\em NuSTAR} mission is designed to have access to 80\% of the sky at any time,
and has ToO capability to enable follow-up of any Type~Ia supernova out to Virgo and any core collapse in the local
group that occurs during the mission life.

A wide variety of programs are being considered to fill the science reserve time.  These include observations of the Sun to 
search for micro flares believed to heat the corona, high signal to noise spectroscopy of bright AGN, surveys and ToO
observations of known magnetars,  mapping non-thermal emission in galaxy clusters, and imaging of nearby starburst
galaxies.   While a number of these programs will be accomplished in the two-year baseline mission, others will
be good candidates for Guest Observer (GO) proposals undertaken in an extended mission.    {\em NuSTAR} has no
consumables, and the expected orbit lifetime is in excess of five years.   The team expects to propose a GO program to 
the NASA Senior Review that will broaden the scientific return and enable broader community participation.

\section{SCIENCE INSTRUMENT IMPLEMENTATION} 

\begin{figure}
\begin{center}
\includegraphics[height=5cm]{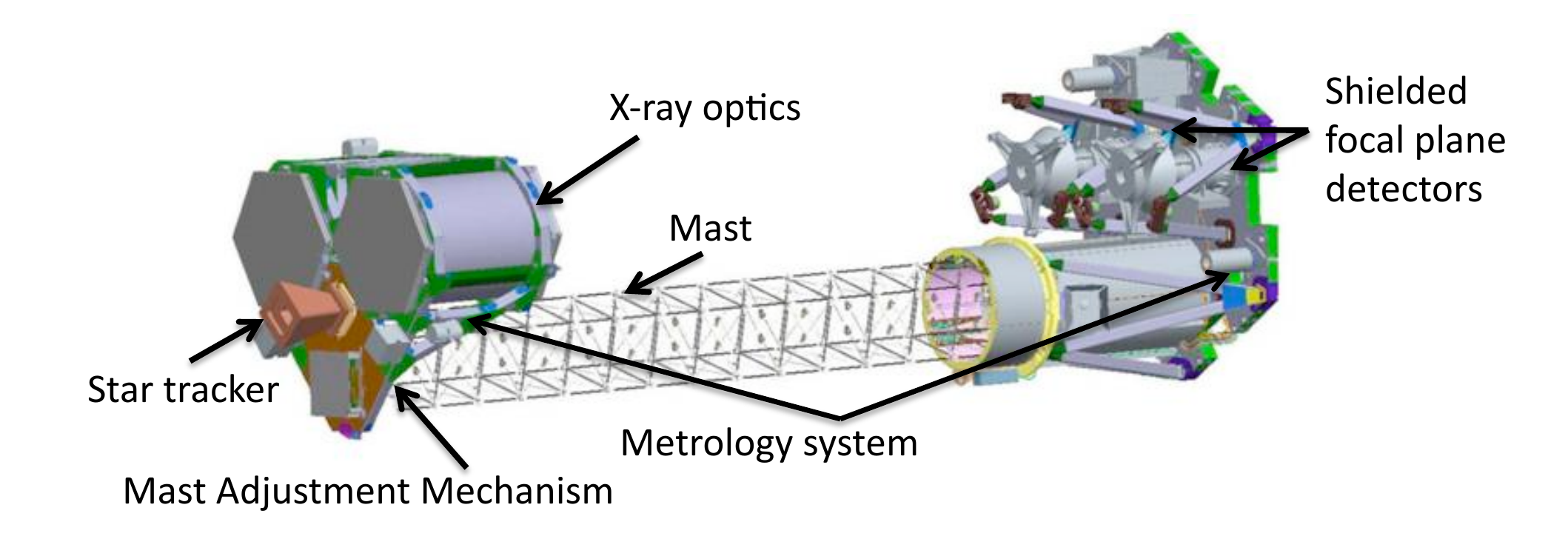}
\end{center}
\caption[inst] 
{\label{fig:inst} 
Diagram of the {\em NuSTAR} instrument showing the principal elements.}
\end{figure} 

The {\em NuSTAR} science instrument (see Figure~\ref{fig:inst}) consists of two co-aligned  grazing incidence optics focusing on to two
shielded solid state CdZnTe pixel detectors.    The instrument is launched in a compact, stowed configuration, and after launch a 
10-meter mast, manufactured by ATK Space Systems, Goleta, is deployed to achieve a focal length of 10.15~m.   
Because the absolute deployment location of the mast is difficult to measure on the ground, due to complications
associated with complete gravity offloading, an adjustment mechanism is built into the last section of the mast to enable a one-time alignment
to optimize the location of the optical axes on the focal plane.   This mechanism provides two angular adjustments as well as rotation.
The mast is not perfectly rigid, but undergoes thermal distortions particularly
when going in and out of Earth shadow (the mission is deployed in a low-Earth orbit) that translate into changes in telescope alignment
of 1 -- 2$^\prime$.   These mast alignment changes are measured by the combination of an optics bench-mounted star tracker and a laser
metrology system.    
The same combination of sensors also provides the absolute instrument aspect.  In order to limit the FoV open
to the detectors, and therefore the diffuse cosmic background, an aperture stop consisting of three rings deploys
with the mast.   The aperture stop is shown in stowed configuration in Figure~\ref{fig:inst}.  In deployed configuration
the top will be 0.83~m above  the focal plane surface.   

\begin{table}[h]
\caption{Optics configuration summary.}
\label{tab:opt}
\begin{center}
\begin{tabular}{|l|c||l|c|}
\hline
{\bf Parameter } & {\bf Value} &  {\bf Parameter } & {\bf Value} \\
\hline\hline
Focal length & 10.15 m & Shell length & 22.5 cm \\ \hline
\# shells & 133 & Min. graze angle & 1.34 mrad \\ \hline
\# azimuthal segments & 6 (inner)/12 ( outer) & Max. graze angle & 4.7 mrad \\ \hline
Inner radius & 5.44 cm & Coating (outer) & W/Si \\ \hline
Outer radius & 19.1 cm & Coating (inner) & Pt/C \\ \hline
\end{tabular}
\end{center}
\end{table}

The {\em NuSTAR} optics utiliThe foe a conical approximation to a Wolter-I design in a highly nested configuration, with 133 shells per optic
with graze angles ranging from 4.6 -- 16 arcminutes.   The shells are fabricated from segmented thermally formed glass, and the segments
are coated with depth-graded multilayers optimized  to achieve significant high energy response  for this graze angle range~\cite{mhm+09}.   
The coatings employ a combination of W/Si bilayers on the outer shells, and Pt/C on the inner shells.   The high-energy 
effective area cutoff results from the K-shell absorption in Platinum.   Hailey~{\em et al.} (2010)~\nocite{hai+10}
describe the optics implementation in detail, and Table~\ref{tab:opt} provides a summary of the primary configuration parameters.

\begin{figure}
\begin{center}
\includegraphics[height=5.5cm]{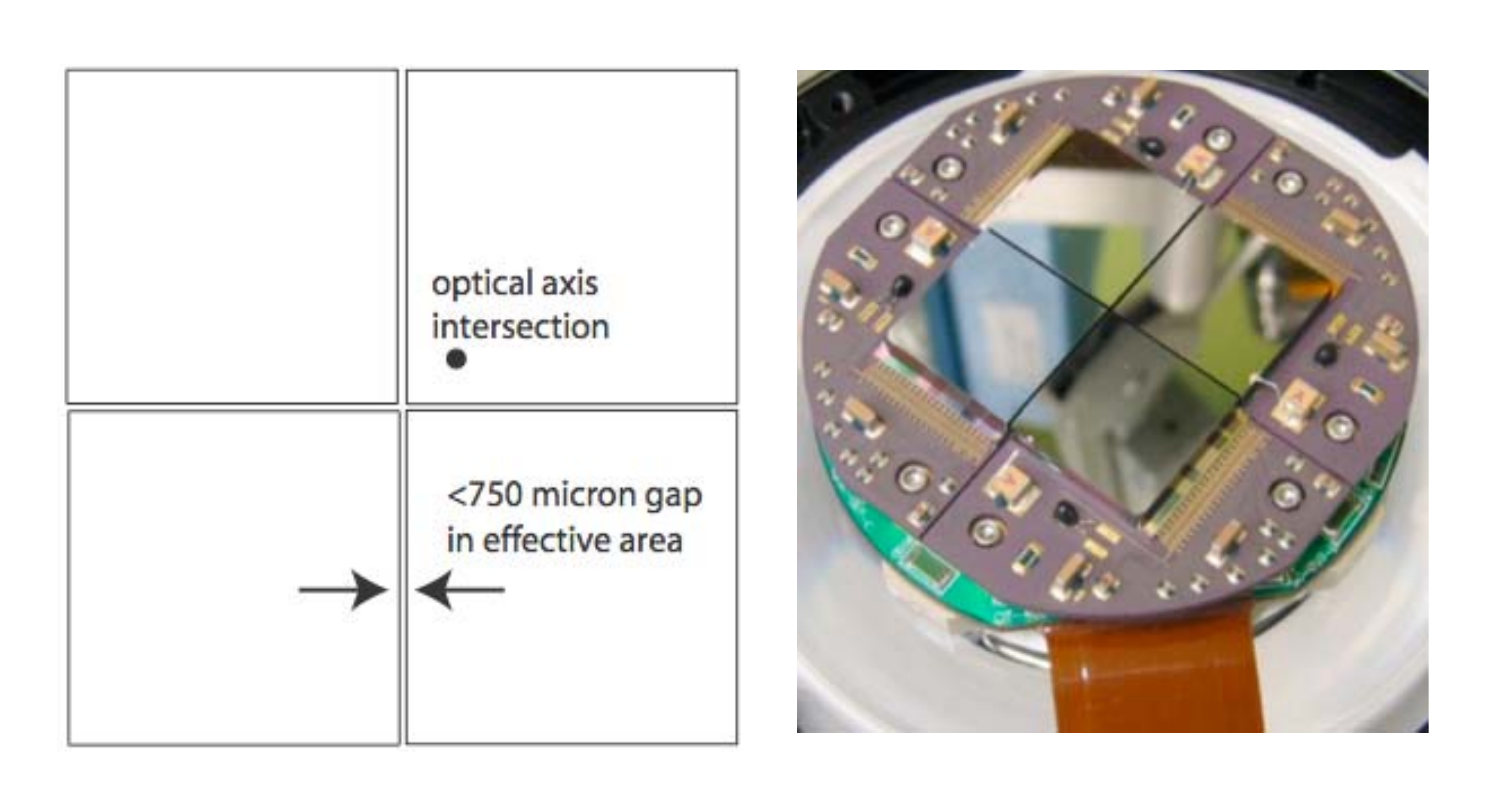}
\end{center}
\caption[fp] 
{\label{fig:fp} 
The {\em NuSTAR} focal plane configuration and photograph of an engineering test module.}
\end{figure}

Each focal plane consists of four CdZnTe pixel sensors coupled to a custom low-noise ASIC~\cite{hcm+10}.  Each hybrid contains a $32 \times 32$ array of
600~$\mu$m pixels with a resulting plate scale of 12.3$^{\prime\prime}$/pixel, so that the mirror point spread function is over-sampled.
The sensors are placed in a two-by-two array with a minimal ($\sim500~\mu$m) gap between them to fill a total subtended field of
view of 13$^\prime$ on a side (Figure~\ref{fig:fp}).    Table~\ref{tab:fp} summarizes the primary characteristics of the focal plane.

\begin{table}[h]
\caption{Focal plane configuration summary.}
\label{tab:fp}
\begin{center}
\begin{tabular}{|l|c||l|c|}
\hline
{\bf Parameter } & {\bf Value} &  {\bf Parameter } & {\bf Value} \\
\hline\hline
Pixel size & 0.6 mm/12.3$^{\prime\prime}$ & Max processing rate & 400 evt/s \\ \hline
Focal plane size & 13$^\prime \times 13^\prime$ & Max flux meas. rate & 10$^4$/s \\ \hline
Pixel format & $32 \times 32 $ &  time resolution &  2$\mu$sec \\ \hline
Threshold & 2.5 keV (each pixel) & Dead time fraction (weak source) & 2\%\\ \hline
\end{tabular}
\end{center}
\end{table}

To achieve a low energy threshold and good spectral performance, the detector readout is designed for very low noise.
The electronic noise contribution (including detector leakage current)
to the energy resolution is 400~eV, and the low-energy threshold is 2.5~keV for an event registering in a single pixel.
Over most of the energy range the detector spectral resolution is limited by charge collection uniformity in the CdZnTe crystal.   At low energies, between
5 and 30 keV,  the average spectral resolution for a typical flight detector is 500~eV FWHM, while at 60~keV it is 1.0~keV, and at 86~keV it is 1.2~keV.
The focal plane will be passively cooled in flight to between 0$^\circ$ and 5$^\circ$~C.  The passive cooling is enabled by the low-power dissipation of the
detector readout chip (50~$\mu$W/pixel).   At in-flight operating temperatures, the detector leakage current is a negligible contributor to the resolution.
In addition to measuring the deposited energy and arrival time for each event, the readout architecture enables a depth of interaction measurement
which can be used both to maximize photo peak efficiency at high energy, where charge trapping effects can lead to a low-energy
`tail' on the energy resolution, and in addition reject background from the back portion of the detector.
   
The readout of each focal plane module is controlled by an FPGA-embedded microprocessor.   Because each pixel triggers independently,
and the electronics shaping time is short, there are no pile-up issues equivalent to the CCD focal planes on {\em XMM} and {\em Chandra}.
The maximum rate that events can be processed is 400~cps in each telescope; however,
pulse pileup does not occur until substantially higher rates ($\sim10^5$ cps).    The readout system is designed so that source fluxes
can be measured up to count rates of $10^4$~cps.    At the nominal faint-source count rates, the readout dead time is $\lsim~2$\%.

\begin{figure}
\begin{center}
\includegraphics[height=7.5cm]{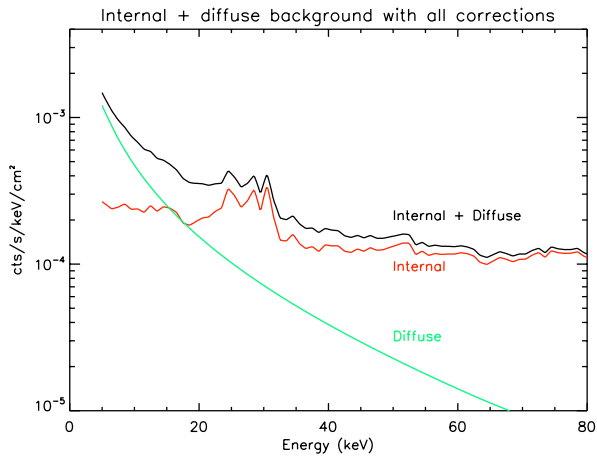}
\end{center}
\caption[bkg] 
{\label{fig:bkg} 
Predicted detector background count rate per unit area as a function of energy.}
\end{figure} 

The focal plane is surrounded by an active 2~cm thick CsI(Na) shield and incorporates a deployable aperture stop.   The CsI shield extends
20~cm above the detector, and has an opening angle of 16~degrees, while the passive aperture stop defines a much narrower opening of
4~degree diameter.  Figure~\ref{fig:bkg} shows the expected background counts per unit detector area as calculated using the GEANT-based
MGGPOD\cite{wsn+07}  Monte Carlo suite.  At low energies the background is 
dominated by diffuse leakage through the portion of the aperture stop FoV not blocked by the optics bench.   The spectral features
between 25 and 35~keV are fluorescence from the CsI shield.   The background level shown in Figure~\ref{fig:bkg}
assumes the use of the depth-of-interaction measurement
to reject interactions in the back of the detector, which results in about a factor two background reduction at 60~keV.

The instrument components are in a mature state, with
flight hardware fabrication underway, and delivery to integration and test beginning in Fall 2010.

\section{MISSION DESIGN}

{\em NuSTAR} will be launched on a Pegasus XL rocket into a 6$^\circ$ inclination, $575 \times 600$~km low Earth orbit.   The low inclination,
achieved by using a near-Equatorial launch site at Kwajalein Island, significantly reduces internal detector backgrounds because the
orbit avoids the intense region of the SAA (which it still skims on some orbits).   The orbit altitude leads to an expected mission life
in excess of five years.   The telescope will not re-orient during Earth occultations, and the total expected observing efficiency
is 50\% for typical targets, with higher efficiency (approaching 90\%) for targets near the poles.  The spacecraft, manufactured by Orbital Sciences Corp.,
is three-axis stabilized with a single articulating solar panel and relies predominantly upon a multi-head star camera (the DTU microASC)
for aspect.   This enables 80\% of the sky to be accessed at any given time, which both allows ToO viewing with few restrictions, and
aids in mission planning.   

The satellite will be operated out of the Mission Operations Center (MOC) at U.C. Berkeley.   Command uplinks and data downlinks will be
through a ground station, operated by the Agenzia Spaziale Italiana, located in Malindi, Kenya.  Most science targets will be viewed for a 
week or more, so that after a 30-day in-orbit checkout and commissioning period, commanding will be rare.   The turnaround time for
ToO observations depends largely on timing relative to the ground station passes.   The Malindi station is visible
once per 90~minute orbit, but commands can take up to 12 hours to prepare given that the MOC is not staffed 24~hours/day.
The spacecraft slews at an average rate of 1.2$^\circ$/minute, so a typical slew will take less than 90~minutes.

\section{GROUND DATA SYSTEM}

The science data will be transferred from the U.C. Berkeley MOC to a Science Operations Center (SOC) located at Caltech.    The SOC
will process and validate the data, and distribute products to the science team.    All science data will be converted to FITS
format conforming to Office of Guest Investigator Programs (OGIP) standards, and analysis software will adopt the FTOOLS approach and environment.
The {\em NuSTAR} science data has no
proprietary period, and after a six-month interval during which the instrument calibration will be understood and the
performance verified, data will enter the public science archive, located at the HEASARC at
Goddard Space Flight Center, within two months of completion of an observation (the two-month period being required for
ground data processing and validation).

\section{SUMMARY AND CONCLUSIONS}

With a launch scheduled for early 2012, {\em NuSTAR} will be the first of a new generation of focusing hard X-ray telescopes, 
and will provide two orders of magnitude improvement in sensitivity over previous hard X-ray missions, combined with
good spectral resolution and sub-arcminute imaging.      In the first two years {\em NuSTAR} will focus on 
four key science objectives.  However, the mission life is limited only by the orbit decay, and an extended mission will be proposed
to allow a substantial guest investigator program to expand the scientific reach.   
The scientific performance expectations are largely based on measurements
made on prototype and flight hardware, giving confidence in current projections for sensitivity, spectroscopic and imaging performance.

{\em NuSTAR} will incorporate a number of recently developed technologies.
At the heart of the instrument are novel depth-graded multilayer optics focusing onto CdZnTe pixel detectors developed originally
for the {\em HEFT} balloon experiment.   {\em NuSTAR}
will be the first X-ray telescope to utilize a long (10-meter)  extendable mast, and will demonstrate the application of this
technology as well as the requisite metrology system.     While the instrument contains new elements, the spacecraft
is based on a heritage design, and the mission operations approach is simple.   These elements enable a highly capable
science mission  on a Small Explorer platform.


\acknowledgments     
 
The {\em NuSTAR} mission is funded by NASA through contract number NNG08FD60C. 
Additional contributions are provided by
the Danish Technical University for optics coating and calibration and the Agenzia Spaziale Italiana  (ASI) for the Malindi ground station and ground
data system development.  Mission management is provided by
the Jet Propulsion Laboratory, and program management provided by the Explorer Program Office at Goddard Space Flight Center.   


\bibliography{SPIE}   
\bibliographystyle{spiebib}   

\end{document}